\documentclass[12pt]{article}

\font\grande=cmr9.5 scaled \magstep4
\font\medio=cmr9.5 scaled \magstep2
\outer\def\beginsection#1\par{\medbreak\bigskip
      \message{#1}\leftline{\bf#1}\nobreak\medskip
\vskip-\parskip
      \noindent}
\usepackage{graphicx} 
\begin{document}
\bibliographystyle {unsrt}

\titlepage

\begin{flushright}
CERN-TH-2016-147
\end{flushright}

\vspace{15mm}
\begin{center}
{\grande Anomalous magnetohydrodynamics}\\
\vspace{5mm}
{\grande in the extreme relativistic domain}\\
\vspace{15mm}
 Massimo Giovannini 
 \footnote{Electronic address: massimo.giovannini@cern.ch} \\
\vspace{0.5cm}
{{\sl Department of Physics, Theory Division, CERN, 1211 Geneva 23, Switzerland }}\\
\vspace{1cm}
{{\sl INFN, Section of Milan-Bicocca, 20126 Milan, Italy}}
\vspace*{1cm}

\end{center}

\centerline{\medio  Abstract}
\vskip 0.5cm
The evolution equations of anomalous magnetohydrodynamics are derived in the extreme relativistic regime and contrasted with the  
treatment of hydromagnetic nonlinearities pioneered by Lichnerowicz in the absence of anomalous currents. 
In particular we explore the situation where the conventional vector currents are complemented by the axial-vector currents 
arising either from the pseudo Nambu-Goldstone bosons of a spontaneously broken symmetry or because 
of finite fermionic density effects. After expanding the generally covariant equations in inverse powers of the conductivity, 
the relativistic analog of the magnetic diffusivity equation is derived in the presence of vortical and magnetic currents. 
While the anomalous contributions are generally suppressed by the diffusivity, they are shown to disappear in the perfectly 
conducting limit. When the flow is irrotational, boost-invariant and with vanishing four-acceleration the corresponding evolution 
equations are explicitly integrated so that the various physical regimes can be directly verified.
\noindent

\vspace{5mm}

\vfill
\newpage
The non-relativistic evolution of hydromagnetic nonlinearities in charged liquids at high magnetic Reynolds numbers
leads to effective currents that are parallel rather than orthogonal to the orientation of the magnetic field. This situation is realized, for instance, 
in the context of turbulent dynamos where the plasma's kinetic energy amplifies the large-scale magnetic field if and when the bulk velocity $\vec{v}$
is incompressible and, in average, non-mirror symmetric i.e. $\langle \vec{v} \cdot \vec{\nabla} \times \vec{v} \rangle \neq 0$ \cite{one}. 
A similar physical system occurs when a globally neutral plasma contains vector and axial-vector 
currents arising either from the effective action of gauge fields at finite fermionic density \cite{two} or from
some pseudo Nambu-Goldstone boson of a spontaneously broken symmetry interacting with fermions \cite{three}.  
Anomalous magnetohydrodynamics (AMHD) aims exactly at describing the dynamical evolution of the gauge fields in a plasma containing both vector and axial-vector currents \cite{four}. The relativistic theory of the ordinary hydromagnetic nonlinearities has been shaped long ago by Lichnerowicz and developed by 
various authors \cite{five}.  It seems therefore both interesting and natural to relax the assumption that the hypermagnetic fields are merely external and to formulate AMHD in the extreme relativistic regime by including on equal footing the axial-vector and Ohmic currents. The obtained results can be relevant for two 
complementary areas namely the dynamics of the magnetized electroweak phase \cite{six} and the hydrodynamic models of multiparticle dynamics \cite{seven,eight}. 

The very notion of relativistic magnetic fields might appear as an oxymoron insofar as the electric and magnetic fields are 
non-relativistic concepts that must be replaced, in a Lorentz covariant formulation,  by the appropriate field strength tensor $Y_{\mu\nu}$ 
(and by its dual\footnote{
The totally antisymmetric symbol of Levi-Civita in four-dimensions is denoted by $\epsilon^{\mu\nu\alpha\beta}$ while 
$E^{\mu\nu\alpha\beta} = \epsilon^{\mu\nu\alpha\beta}/\sqrt{-g}$ (where $g = \mathrm{det}\,g_{\mu\nu}$) 
transforms correctly as a contravariant tensor under general coordinate transformations. Note that $g_{\mu\nu}$ denotes the metric tensor (with signature $(+,\, -,\,-,\, -)$) of a four-dimensional space-time geometry.  Units $\hbar=c=1$ will be used throughout (so that 
where, for instance,  $\gamma= 1/\sqrt{ 1 - v^2}$).} $\widetilde{Y}^{\mu\nu} = E^{\mu\nu\alpha\beta} Y_{\alpha\beta}/2$).  However, if there exist a family of four-dimensional observers moving with four-velocity $u^{\mu}$ the relativistic dynamics of hydromagnetic nonlinearities can be described in terms of two generalized electric and magnetic fields defined, respectively, as \cite{five}:
\begin{equation}
Y_{\mu\nu} = {\mathcal E}_{[\mu}\, u_{\nu]}\, + \, E_{\mu\nu\rho\sigma}\, u^{\rho} {\mathcal B}^{\sigma},\qquad 
\widetilde{Y}^{\mu\nu} = {\mathcal B}^{[\mu}\, u^{\nu]}\, + \, E^{\mu\nu\rho\sigma}\, {\mathcal E}_{\rho} u_{\sigma},
\label{eq1}
\end{equation}
where  $ {\mathcal E}_{[\mu}\, u_{\nu]} = {\mathcal E}_{\mu} u_{\nu} - {\mathcal E}_{\nu} u_{\mu}$ (and similarly for ${\mathcal B}^{[\mu}\, u^{\nu]}$); ${\mathcal E}^{\mu} = Y^{\mu\nu}\, u_{\nu}$ and 
${\mathcal B}^{\mu} = \widetilde{Y}^{\mu\nu}\, u_{\nu}$  generalize the electric and the magnetic components to the relativistic regime. Bearing in mind that $Y^{i0} = e^{i}$ and $Y^{ij} = - \epsilon^{ijk} b_{k}$, Eq. (\ref{eq1}) implies that the two four-vectors can also be written, in three-dimensional notation and in flat space-time, as ${\mathcal E}^{\mu} = \gamma( \vec{e}\cdot\vec{v}, \,\, \vec{e} + \vec{v} \times \vec{b})$ and ${\mathcal B}^{\mu} = \gamma(- \vec{b}\cdot\vec{v}, \,\, - \vec{b} + \vec{v} \times \vec{e})$.
The vector current $j_{\alpha}$  couples to the hypercharge field and it is not anomalous so that this part of the model describes an unbroken $U(1)$ gauge theory \cite{four}. Since we want the gauge fields to be dynamical, the total action of the problem can be written as:  
\begin{eqnarray}
S_{total} &=& \int d^{4} x \, \sqrt{-g} \biggl\{ - \frac{R}{16 \pi G} + \frac{1}{2} g^{\alpha\beta} \partial_{\alpha} \psi\, \partial_{\beta} \psi - W(\psi) 
- \frac{1}{16 \pi}  Y_{\alpha\beta} \, Y^{\alpha\beta} 
\nonumber\\
&-&j_{\alpha} \, Y^{\alpha} - \frac{1}{16 \pi}\biggl[ \alpha_{B} \frac{\psi}{M} Y_{\alpha\beta} \widetilde{Y}^{\alpha\beta}+ 2 \frac{\alpha_{\omega}}{8 \pi} \, \psi\, Y_{\alpha\beta}\, \widetilde{\omega}^{\alpha\beta}\biggr]  + \,.\,.\,.\biggr\} + S_{m},
\label{eq2}
\end{eqnarray}
where $R$ is the Ricci scalar and $G$ is the Newton constant; $S_{m}$ denotes the matter part of the action (taken to be in a perfect fluid form) while the ellipses stand for further interactions\footnote{The ellipses in Eq. (\ref{eq2}) may stand for further terms of the type
 $\omega_{\beta} {\mathcal E}^{\beta}$, $\psi\, \omega_{\alpha} {\mathcal B}^{\alpha}$ (and so on and so forth) where $\omega^{\alpha}$ is the vorticity four-vector (see Eq. (\ref{eq3})). These terms will be neglected but they can be easily included.}. In Eq. (\ref{eq2})  $\psi$ denotes the pseudo Nambu-Goldstone field characterized by the potential $W(\psi)$ and symmetry breaking scale $M$; the two dimensionless constants $\alpha_{B}$ and $\alpha_{\omega}$ parametrize, respectively, the couplings of $\psi$ with the gauge field and with the vorticity of the fluid. The vorticity four-vector is defined as  $\omega^{\alpha} = \widetilde{\omega}^{\alpha\beta}\, u_{\beta}$ and we assume the conventional decomposition of the generally covariant derivative\footnote{ Inverting Eq. (\ref{eq3}) 
 we can also write $u_{\alpha\,; \beta} = \dot{u}_{\alpha} \, u_{\beta} \, + \sigma_{\alpha\beta}\, + \omega_{\alpha\beta} +  \theta {\mathcal P}_{\alpha\beta}/3$ which is 
 the standard decomposition of the covariant derivative.}:
\begin{eqnarray}
\omega_{\alpha\beta} = \frac{1}{2}u_{[\alpha\, ; \beta]}  - \frac{1}{2}\dot{u}_{[\alpha} \, u_{\beta]} ,\qquad
\sigma_{\alpha\beta} = \frac{1}{2}(u_{\alpha\, ; \beta} + u_{\beta\, ; \alpha}) - \frac{1}{2}(\dot{u}_{\alpha} \, u_{\beta} + \dot{u}_{\beta} u_{\alpha}) - \frac{\theta}{3} {\mathcal P}_{\alpha\beta},
\label{eq3}
\end{eqnarray}
 where $\sigma_{\alpha\beta}$ is the shear tensor and $\theta = \nabla_{\alpha} u^{\alpha}$. As usual the semicolon stands for the covariant derivative (i.e. $u_{\alpha\, ;\beta} = \nabla_{\beta} u_{\alpha}$) while the overdot  denotes the absolute derivative in the direction of $u^{\gamma}$ (i.e. $\dot{u}_{\alpha} = u^{\gamma} \nabla_{\gamma} u_{\alpha}$); finally  ${\mathcal P}_{\alpha\beta} = g_{\alpha\beta} - u_{\alpha} u_{\beta}$ is the standard covariant projector. 

Since the variation of $S_{m}$ in Eq. (\ref{eq2}) leads to the energy-momentum tensor of a perfect fluid (i.e. $T_{\mu\nu} = \rho u_{\mu} u_{\nu} - p\, {\mathcal P}_{\mu\nu}$),  the three relevant evolution equations inferred from Eq. (\ref{eq2}) can be written as:
\begin{eqnarray}
&& g^{\alpha\beta} \nabla_{\alpha} \nabla_{\beta}\psi + W_{,\,\psi}= - \frac{\alpha_{B}}{16\pi M} Y_{\alpha\beta} \, \widetilde{Y}^{\alpha\beta} - \frac{\alpha_{\omega}}{8 \pi} Y_{\alpha\beta}\, \widetilde{\omega}^{\alpha\beta},
\label{eq4}\\
&& \nabla_{\alpha} Y^{\alpha\beta} = 4 \pi j^{\beta} - \frac{\alpha_{B}}{M}\partial_{\alpha} \psi \widetilde{Y}^{\alpha\beta} - \alpha_{\omega} \biggl[(\partial_{\alpha} \psi)\,\widetilde{\omega}^{\alpha\beta} + \psi \nabla_{\alpha} \widetilde{\omega}^{\alpha\beta}\biggr],
\label{eq5}\\
&& \nabla_{\mu} T^{\mu}_{\nu} = Y_{\nu\alpha} j^{\alpha} - \frac{\alpha_{\omega}}{4\pi} \biggl[ (\partial_{\mu}\psi)\, \widetilde{\omega}^{\mu\alpha} + \psi \nabla_{\mu} \widetilde{\omega}^{\mu\alpha}\biggr] Y_{\nu\alpha} + \frac{\alpha_{\omega}}{8 \pi} (\partial_{\nu}\psi) \, Y_{\alpha\beta} \, \widetilde{\omega}^{\alpha\beta},
\label{eq6}
\end{eqnarray}
where the notation $W_{,\,\psi} =\partial W/\partial \psi$ has been used;
the Bianchi identity for the gauge field implies that $ \nabla_{\alpha} \widetilde{Y}^{\alpha\beta} =0$ while  $\nabla_{\alpha} \widetilde{\omega}^{\alpha\beta} = - \nabla_{\alpha} [ E^{\alpha\beta\rho\sigma} \dot{u}_{\rho} u_{\sigma}]$ so that $\nabla_{\alpha} \widetilde{\omega}^{\alpha\beta}=0$  only when the four-acceleration of the fluid flow vanishes (i.e. $\dot{u}_{\rho}=0$).  Of course Eqs. (\ref{eq4}) and (\ref{eq5}) can also be phrased in terms of the corresponding energy-momentum tensors:
\begin{eqnarray}
\nabla_{\mu} {\mathcal S}^{\mu}_{\nu} &=& - \frac{\alpha_{B}}{16 \pi M} \,\partial_{\nu} \psi \,Y_{\alpha\beta} \widetilde{Y}^{\alpha\beta} - \frac{\alpha_{\omega}}{8\pi} \,\partial_{\nu} \psi \,Y_{\alpha\beta} \widetilde{\omega}^{\alpha\beta},
\label{eq7}\\
\nabla_{\mu} {\mathcal T}^{\mu}_{\nu} &=& - Y_{\nu\alpha}\, j^{\alpha}  + \frac{\alpha_{\omega}}{4\pi}\biggl[ (\partial_{\mu}\psi) \, \widetilde{\omega}^{\mu\alpha} + \psi \nabla_{\mu} \widetilde{\omega}^{\mu\alpha} \biggr]  Y_{\nu\alpha} + \frac{\alpha_{B}}{16 \pi M} \, (\partial_{\nu} \psi)\, Y_{\alpha\beta} \, \widetilde{Y}^{\alpha\beta},
\label{eq8}
\end{eqnarray}
where ${\mathcal S}^{\mu}_{\nu}$ and ${\mathcal T}^{\mu}_{\nu}$ are given, respectively, by:
\begin{eqnarray}
{\mathcal S}_{\nu}^{\mu} &=& \partial_{\nu} \psi \partial^{\mu} \psi - \delta_{\mu}^{\nu} \biggl( \frac{1}{2} g^{\alpha\beta} \partial_{\alpha} \psi \partial_{\beta} \psi - W\biggr),
\label{eq9}\\
{\mathcal T}_{\nu}^{\mu} &=&  \frac{1}{4 \pi} \biggl[ -\biggl( Y_{\nu\alpha}\, Y^{\mu\alpha} + \alpha_{B}\frac{\psi}{M} \, Y_{\nu\alpha}\, \widetilde{Y}^{\mu\alpha} \biggr) 
 + \frac{1}{4} \delta_{\nu}^{\mu} 
 \biggl( Y_{\alpha\beta} \, Y^{\alpha\beta} + \alpha_{B} \frac{\psi}{M} Y_{\alpha\beta} \, \widetilde{Y}^{\alpha\beta}\biggr) \biggr].
\label{eq10}
\end{eqnarray}
By  summing up Eqs. (\ref{eq6}), (\ref{eq7}) and (\ref{eq8}) it can be easily verified that the total energy-momentum tensor $T_{\mu\nu}^{(tot)} = T_{\mu\nu} + {\mathcal S}_{\mu\nu} + {\mathcal T}_{\mu\nu}$ 
is covariantly conserved (i.e. $\nabla_{\mu} T^{\mu\nu}_{(tot)} =0$) as implied by the Bianchi identity of the corresponding Einstein 
equations $(R_{\mu\nu} - R g_{\mu\nu}/2)= 8 \pi G \,T_{\mu\nu}^{(tot)}$ where $R_{\mu\nu}$ is the Ricci tensor and $R$ is the Ricci scalar. 
The covariant conservation of the total energy-momentum tensor implies the conservation of the total entropy four-vector\footnote{If only  some selected parts of the system are taken into account the entropy of the subsystem may appear to be not conserved. This is, in particular, what happens if the gauge fields are considered as external sources.}. When the anomalous charge is not conserved because of the Abelian anomaly (i.e.  $\nabla_{\mu} j^{\mu}_{R} = - g^{\prime\, 2}\,Y_{\alpha\beta} \, \widetilde{Y}^{\alpha\beta}/(64 \pi^2)$ where $g^{\prime}$ is the hypercharge coupling) the effective action for the Abelian fields at finite fermionic density can be written as \cite{two}:
\begin{equation}
S= - \frac{g^{\prime\, 2}}{64 \pi^3} \int d^{4} x \sqrt{-g} \, \mu_{R} \, E^{\mu\nu\alpha\beta} \,Y_{\mu\nu} \, Y_{\alpha} \, v_{\beta},\qquad g^{\alpha\beta} v_{\alpha} v_{\beta} =1,
\label{eq12}
\end{equation}
showing that the modulus of the covariant derivative of the pseudo Nambu-Goldstone boson coincides with the chemical potential\footnote{Equation (\ref{eq12}) accounts for the same interaction between $\psi$ and $Y_{\alpha\beta} \widetilde{Y}^{\alpha\beta}$ provided we identify $\alpha_{B} = \alpha_{Y}/\pi$ and $\partial_{\mu}\psi = M \, \mu_{R} \, v_{\mu}$ (with $\alpha_{Y} = g^{\prime \, 2}/(4\pi)$ and $\mu_{R} = \sqrt{g^{\alpha\beta}\partial_{\alpha}\psi\partial_{\beta}\psi}/M$).}.

In the case of a highly conducting plasma the vector current appearing in Eqs. (\ref{eq5}) and (\ref{eq8}) can be written as  the sum of two terms, namely $j^{\mu} = \sigma_{c} \, {\mathcal E}^{\mu}  + n \, u^{\mu}$ where $\sigma_{c}$ is the conductivity (potentially very large) and $n= j^{\mu}\, u_{\mu} $ is the charge concentration. In analogy with the relativistic treatment of hydromagnetic nonlinearities the relevant equations shall be expanded in inverse powers of the conductivity or, more formally, in powers of a dimensionless parameter $\varepsilon$ defined as:
\begin{equation}
\varepsilon = \frac{1}{4 \,\pi\,  L \, \sigma_{c}}  = \frac{\eta}{L} \simeq \eta \nabla_{\alpha}, \qquad \eta= \frac{1}{4\pi \sigma_{c}},
\label{eq13}
\end{equation}
where $\eta$ denotes the magnetic diffusivity\footnote{Even if $\eta$ often stands for 
the (pseudo)rapidity, we shall denote the magnetic diffusivity by $\eta$ (as it is traditional in plasma literature); the rapidity will be denoted by $y$ as in hydrodynamical models of multiparticle collisions \cite{seven}.} and $L$ is the typical scale of variation of the covariant gradients.  After expanding all the dynamical equations in powers of $\varepsilon$ the fate of the anomalous contributions can either be studied in the ideal limit (coinciding with the perfectly conducting regime where $\varepsilon\to 0$) or in the resistive approximation where $\sigma_{c}$ may be very large remaining however always finite (i.e. $\varepsilon <1$). The first step will be to express the Bianchi identity $\nabla_{\alpha} \widetilde{Y}^{\alpha\beta}=0$ in terms of Eq. (\ref{eq1});  the result of this exercise will give  
$\nabla_{\alpha} {\mathcal B}^{[\alpha} u^{\beta]}+ E^{\alpha\beta\rho\sigma} \nabla_{\alpha} ({\mathcal E}_{\rho}\, u_{\sigma}) =0$. The previous 
equation can then be projected along $u_{\beta}$ and the final result becomes:
\begin{equation}
\nabla_{\alpha} {\mathcal B}^{\alpha} + {\mathcal B}^{\beta} \, \dot{u}_{\beta} = 2  \omega_{\rho} \, {\mathcal E}^{\rho} \to 8\pi \,\eta \,\omega_{\rho} j^{\rho},
\label{eq15}
\end{equation}
where the right hand side follows from ${\mathcal B}^{\beta} \dot{u}_{\beta} = - \dot{{\mathcal B}}^{\beta} u_{\beta}$ (since, by definition, $\nabla_{\alpha} (u_{\beta} {\mathcal B}^{\beta}) =0$). In Eq. (\ref{eq15}) the expression preceded by an arrow is derived by trading in $2 \omega_{\rho} {\mathcal E}^{\rho}$ the hyperelectric field for the current, i.e. ${\mathcal E}^{\rho}  = (j^{\rho} - n u^{\rho})/\sigma_{c}$; recall, in this respect, that $\omega_{\rho} u^{\rho} =0$. The same procedure  can be applied to Eq. (\ref{eq4}) and this time the result is:
\begin{equation}
g^{\alpha\beta} \nabla_{\alpha} \nabla_{\beta}\psi + W_{,\, \psi} + \frac{\alpha_{\omega}}{8 \pi} E_{\alpha\beta\rho\sigma} u^{\rho} {\mathcal B}^{\sigma}\, \widetilde{\omega}^{\alpha\beta}  = - \frac{\alpha_{B}}{4\pi M} {\mathcal E}_{\alpha} {\mathcal B}^{\alpha}  - \frac{\alpha_{\omega}}{4 \pi} {\mathcal E}_{\alpha}\, u_{\beta}\, \widetilde{\omega}^{\alpha\beta}.
 \label{eq16}
 \end{equation}
The terms containing the electric fields have been collected at the right hand side of Eq. (\ref{eq16}) since they are subleading in the 
conductivity expansion.  With the same logic and with the same notations Eq. (\ref{eq5}) can be rewritten as: 
 \begin{eqnarray}
 && \nabla_{\alpha} ( u_{\rho} {\mathcal B}_{\sigma}) E^{\alpha\beta\rho\sigma} - 4\pi j^{\beta} +
\alpha_{\omega} \biggl(\partial_{\alpha}\psi \, \widetilde{\omega}^{\alpha\beta} + \psi \nabla_{\alpha} \widetilde{\omega}^{\alpha\beta}\biggr) + 
\frac{\alpha_{B}}{M} \partial_{\alpha} \psi {\mathcal B}^{[\alpha} u^{\beta]}  =
\nonumber\\  
&& - \nabla_{\alpha} {\mathcal E}^{[\alpha} u^{\beta]} 
 - \frac{\alpha_{B}}{M} \partial_{\alpha} \psi  E^{\alpha\beta\rho\sigma} {\mathcal E}_{\rho} u_{\sigma}.
\label{eq17}
\end{eqnarray}
By projecting Eq. (\ref{eq17}) along $u_{\beta}$ we obtain:
\begin{equation}
\nabla_{\alpha} {\mathcal E}^{\alpha} + \dot{u}^{\beta} {\mathcal E}_{\beta} = 4 \pi j^{\beta} u_{\beta} - \alpha_{\omega}\psi u_{\beta} \nabla_{\alpha} \widetilde{\omega}^{\alpha\beta} + \biggl[\omega_{\alpha} - 
\frac{\alpha_{B}}{M} \partial_{\alpha}\psi\biggr] {\mathcal B}^{\alpha} + \biggl[ {\mathcal B}_{\alpha} - \alpha_{\omega} \partial_{\alpha} \psi\biggr]\omega^{\alpha}.
\label{eq18}
\end{equation}
In the concrete examples discussed hereunder the condition $j^{\beta} u_{\beta} =0$ shall be assumed. However 
all the equations are generally applicable also when the global charge concentration does not vanish.

To lowest order in the $\varepsilon$ expansion, Eq. (\ref{eq17}) should be viewed as an explicit expression for the total current while the remaining 
terms (containing more insertions of hyperlectric fields in various combinations) are irrelevant for the present purposes but they
can be easily determined by going to higher orders in the $\varepsilon$-expansion\footnote{Equation (\ref{eq19}) implies that 
the corrections to the vector current are already ${\mathcal O}(\varepsilon)$. The induced hyperelectric fields will be a fortiori negligible since they turn out to be, to lowest order, ${\mathcal O}(\varepsilon^2)$. }:
\begin{equation}
j^{\beta} = \frac{1}{4 \pi} \nabla_{\lambda} [ E^{\lambda\beta\gamma\delta} u_{\gamma} {\mathcal B}_{\delta} ] + \frac{\alpha_{B}}{4 \pi M} \partial_{\lambda}\psi 
{\mathcal B}^{[\lambda} u^{\beta]} + 
\frac{\alpha_{\omega}}{4 \pi}[ \partial_{\lambda} \psi \widetilde{\omega}^{\lambda\beta} + \psi \nabla_{\lambda} \widetilde{\omega}^{\lambda\beta}] + {\mathcal O}(\varepsilon).
\label{eq19}
\end{equation} 
Equation (\ref{eq19}) shall now be substituted back into the Bianchi identity $\nabla_{\alpha} \widetilde{Y}^{\alpha\beta}=0$ and the final result will be the wanted generalization of the hypermagnetic diffusivity equation  written in a generally covariant language and in the presence 
of anomalous currents:
\begin{eqnarray}
&& \nabla_{\mu}{\mathcal B}^{[\mu} u^{\nu]}- \nabla_{\mu}[\eta\, E^{\sigma\mu\nu\rho} E_{\gamma\delta\lambda\rho} u_{\sigma}
\nabla^{\lambda} (u^{\gamma} {\mathcal B}^{\delta})]  - \frac{\alpha_{B}}{M} \nabla_{\mu}[ \eta \,E^{\mu\nu\rho\sigma}\, {\mathcal B}_{\rho} \dot{\psi}\, u_{\sigma}]
\nonumber\\
&&
+ \alpha_{\omega} \nabla_{\mu} \{\eta \,u_{\sigma}[E^{\mu\nu\rho\sigma}\, (\partial^{\lambda}\psi) \, \widetilde{\omega}_{\lambda\rho} + \psi \nabla^{\lambda} \widetilde{\omega}_{\lambda\rho} ]\}  +{\mathcal O}(\varepsilon^2)=0.
\label{eq21}
\end{eqnarray}
 The first term of Eq. (\ref{eq21}) does not contain any power of the diffusivity; consequently this is the only term surviving 
in the perfectly conducting limit. The second term of Eq. (\ref{eq21}) represents the standard magnetic diffusivity contribution. The two remaining 
contributions correspond to the hypermagnetic and to the vortical currents. To this order in the expansion the hyperelectric fields of Eqs. (\ref{eq15})--(\ref{eq16}) and (\ref{eq17}) are neglected but can be relevant to higher order in the expansion or in the situations where the conductivity is minute.
Using the properties of the Levi-Civita symbols and making explicit their contractions Eq. (\ref{eq21}) becomes:
\begin{eqnarray}
&& \nabla_{\mu}{\mathcal B}^{[\mu} u^{\nu]}  - \nabla_{\mu}\biggl[\eta\, \biggl(\nabla^{[\mu} {\mathcal B}^{\nu]} 
+ u_{\alpha} u^{[\mu} \nabla^{\nu]} {\mathcal B}^{\alpha} +  {\mathcal B}^{[\mu}\dot{u}^{\nu]} + \dot{{\mathcal B}}^{[\mu} u^{\nu]} \biggr)\biggr]
\nonumber\\
&&
 -  \frac{\alpha_{B}}{M} \nabla_{\mu}\biggl( \eta \,E^{\mu\nu\rho\sigma}\, {\mathcal B}_{\rho} \dot{\psi}\, u_{\sigma}\biggr)
+ \alpha_{\omega} \nabla_{\mu} \biggl\{\eta \,u_{\sigma}\biggl[E^{\mu\nu\rho\sigma}\, (\partial^{\lambda}\psi) \, \widetilde{\omega}_{\lambda\rho} + \psi \nabla^{\lambda} \widetilde{\omega}_{\lambda\rho} \biggr]\biggr\}  =0.
\label{eq22}
\end{eqnarray}
If we take the formal limit $\eta \to 0$ (i.e. $\sigma_{c} \to \infty$) in Eq. (\ref{eq22}) 
the only term that survives is the first one. This means that perfectly conducting limit of AMHD in the 
extreme relativistic regime coincides with the perfectly conducting limit in the absence of the anomalous interactions. 
The second term of Eq. (\ref{eq22}) contains two covariant derivatives whereas the chiral magnetic and the 
chiral vortical terms only contain one covariant derivative. Depending on the dynamics of  the anomalous contributions Eq. (\ref{eq22}) suggests that magnetic fields can be amplified in the extreme relativistic limit but the overall result will be anyway scaled down by the initial conductivity of the plasma. 
This qualitative expectation will now be corroborated by a more quantitative discussion since, after all, the magnetic diffusivity may have some 
specific dynamical evolution. 

Even if the conclusions inferred from Eq. (\ref{eq22}) are general (i.e. they do not assume any special profile) we shall now focus the attention, for simplicity,  on the situation where the fluid is 
not vortical (i.e. $\omega_{\alpha\beta} =0$) and the four-acceleration vanishes (i.e. $\dot{u}_{\alpha} =0$). Neglecting the standard magnetic diffusivity term 
(which contains two covariant gradients and which will be anyway included later on when treating a specific class of solutions),  Eq. (\ref{eq22}) becomes
\begin{equation}
\nabla_{\mu}{\mathcal B}^{[\mu} u^{\nu]} -  \alpha_{B} \nabla_{\mu}( \eta \,E^{\mu\nu\rho\sigma}\, {\mathcal B}_{\rho} \dot{\psi}\, u_{\sigma})/M=0.
 \label{eq24}
\end{equation}
Multiplying now both sides of Eq. (\ref{eq24})  by ${\mathcal B}_{\nu}$ and using the covariant decomposition of Eq. (\ref{eq2}) the 
following equation governs the evolution of the magnetic energy density:
\begin{equation}
u^{\alpha}\nabla_{\alpha} B^2+ \frac{4}{3}\theta B^2 + 2 {\mathcal B}^{\mu} {\mathcal B}^{\nu} \sigma_{\mu\nu} = 2 \frac{\alpha_{B}}{M} \eta\, \dot{\psi} E^{\mu\nu\rho\sigma}  (\nabla_{\mu} {\mathcal B}_{\rho})  {\mathcal B}_{\nu}\, u_{\sigma},
\label{eq26}
\end{equation}
where $B^2 = - {\mathcal B}_{\alpha} {\mathcal B}^{\alpha}$ (note that, according to Eq. (\ref{eq1}) ${\mathcal B}_{\alpha} {\mathcal B}^{\alpha} \to- b^2$ in the non-relativistic limit).  The term at the right hand side of Eq. (\ref{eq26}) does not vanish provided 
$E^{\mu\nu\rho\sigma} (\nabla_{\mu} {\mathcal B}_{\rho}) u_{\sigma}$ is proportional to ${\mathcal B}^{\nu}$ either though some 
 constant or through some space-time scalar (such as $\theta$). These conditions generalize, in some sense,  the non-relativistic notion
 of Beltrami fields\footnote{The Beltrami fields are the eigenvalues of the curl operator i.e. $\vec{\nabla}\times \vec{b} = k\, \vec{b}$ where 
 $k$ has dimensions of an inverse length and denotes the typical scale of variation of the magnetic gyrotropy (i.e. $\vec{b} \cdot\vec{\nabla} \times \vec{b}$) in units of the magnetic energy density \cite{one}).} and this is why these constraints shall be referred to as Beltrami conditions. 

The Beltrami constraints are necessary but not sufficient. It is then interesting to solve explicitly the obtained system of equations.  Equations (\ref{eq15}) and 
(\ref{eq26}) demand, respectively, that $\nabla_{\alpha} {\mathcal B}^{\alpha}= - \dot{u}_{\alpha} {\mathcal B}_{\alpha}$ and that $E^{\mu\nu\rho\sigma}  (\nabla_{\mu} {\mathcal B}_{\rho})  {\mathcal B}_{\nu}\, u_{\sigma}\neq 0$.
Both requirements are satisfied at once by the two-dimensional flow $u^{\mu} = [u^{0}(t,x^{3}), \, 0, \, 0,\, u^{3}(t,x^{3})]$ with a
 magnetic field polarized in the orthogonal direction, i.e. ${\mathcal B}^{\mu} =[0, \, {\mathcal B}^{\,1}(t, x^{3}), \, {\mathcal B}^{\,2}(t, x^{3}), \, 0]$. 
 Recalling that $g^{\mu\nu}\,u_{\mu} \,u_{\nu}=1$ and $u_{\alpha} {\mathcal B}^{\alpha}=0$ and requiring that the four-acceleration vanishes (i.e. $\dot{u}_{\mu}=0$) we have that the simplest form of the ansatz is given by:
\begin{equation}
u^{\mu} = (\cosh{y},\, 0,\, 0,\, \sinh{y}), \qquad {\mathcal B}^{\mu} =[0, \, {\mathcal B}^{\,1}(\tau,\, y), \, {\mathcal B}^{\,2}(\tau, \,y ), \, 0],
\label{ex1}
\end{equation}
where the new variables $(\tau, \,y)$ are related to $(t, \, x^{3})$ as $t = \tau \sinh{y}$ and $x^{3} = \tau \cosh{y}$. The result of Eq.  (\ref{ex1}) 
coincides with the boost-invariant flow believed to describe the central rapidity region in the hydrodynamical models 
of multiparticle collisions (see, in particular, third paper of Ref. \cite{seven}).
Defining the two combinations ${\mathcal B}_{\pm} = {\mathcal B}^{1} \pm i  {\mathcal B}^{2}$,
the complete form of Eq. (\ref{eq22}) implies, in Minkowski space-time:
 \begin{equation}
\dot{{\mathcal B}}_{\pm} + \frac{{\mathcal B}_{\pm}}{\tau} =  \frac{\eta^{\prime} {\mathcal B}_{\pm}^{\prime}}{\tau^2} + \frac{{\mathcal B}_{\pm}^{\prime\prime}}{\tau^2} \eta \pm i \alpha_{B} \frac{[\eta\, \dot{\psi}\, {\mathcal B}_{\pm}]^{\prime}}{M \tau},
\label{pmexplicit}
\end{equation}
where the overdot denotes the derivative\footnote{Note that in the case of the irrotational flow of Eq. (\ref{ex1}) the 
absolute derivative coincides with the derivative with respect to $\tau$; notice also that the Minkowski metric must be appropriately used to raise and lower the indices, i.e. ${\mathcal B}_{1} \pm i  {\mathcal B}_{2} = - {\mathcal B}_{\pm}$.} with respect to $\tau$ while the prime denotes a derivation with respect to $y$. Finally, to lowest order in the conductivity expansion we can consistently posit  $\psi = \psi(\tau)$  and $\rho = \rho(\tau)$; thus the profile of Eq. (\ref{ex1}) implies the validity of the following pair of equations
\begin{equation}
\ddot{\psi} + \dot{\psi}/\tau + W_{,\psi}=0, \qquad \dot{\rho} + (\rho + p)/\tau =0,
\label{pmexplicit0}
\end{equation}
coming, respectively, from Eqs. (\ref{eq4}) and (\ref{eq6}) in the case of an irrotational flow with vanishing four-acceleration.
Equations (\ref{pmexplicit}) and (\ref{pmexplicit0}) describe the amplification of the relativistic magnetic field in a direction orthogonal to the flow. However the value of the amplified field does depend predominantly on the initial value of the magnetic diffusivity, as we shall now show. To solve Eq. (\ref{pmexplicit}) we note that $\eta = \eta(\tau, y)$ implying that, in general terms, $\eta^{\prime} \neq 0$. The variation of the conductivity in rapidity leads to the formation of a coherent magnetic with a mechanism that could be viewed as the relativistic analog of the situation arising in the  case of terrestrial dynamos where the magnetic diffusivity has either a radial dependence or is even allowed to fluctuate in space \cite{nine}. Even if this situation is per se interesting it is not directly central to the dynamics of the anomalous currents. We shall then stick to the case  $\eta^{\prime}=0$ which is incidentally the most reasonable if local thermal equilibrium is posited, as we shall specify in a moment.

The physically interesting solution of Eq. (\ref{pmexplicit}) corresponds
to ${\mathcal B}_{\pm}^{\prime}\to 0$ (at least initially) and in the central rapidity region (say for $|y|<1$). If the plasma is locally thermalized $\rho$, $\psi$ and the temperature $T$ will only depend on $\tau$ as Eq. (\ref{pmexplicit0}) implies. The solution of Eq. (\ref{pmexplicit}) will then be:
\begin{eqnarray}
{\mathcal B}_{\pm}(\tau, y) &=& \frac{B_{i}}{\sqrt{Q(\tau)}}\, \biggl(\frac{\tau_{i}}{\tau}\biggr) \, e^{- \frac{y^2 -P^2(\tau)}{4 Q(\tau)} \pm \frac{ i P(\tau) y}{2 Q(\tau)}},
\qquad  {\mathcal B}_{\pm}(\tau_{i}, 0) = B_{i}, \qquad {\mathcal B}_{\pm}^{\prime}(\tau_{i}, 0) =0,
\label{exsol1}\\
P(\tau) &=& - \frac{\alpha_{B}}{M} \int_{\tau_{i}}^{\tau} \frac{\eta(w)}{w} \biggl(\frac{\partial \psi}{\partial w} \biggr) \, dw, \qquad 
Q(\tau) = 1 + \int_{\tau_{i}}^{\tau} \frac{\eta(w)}{w^2} d w,
\label{exsol2}
\end{eqnarray}
where the initial data on $P(\tau)$ and $Q(\tau)$ follow from the boundary conditions of Eq. (\ref{exsol1}). The initial value of the magnetic field at $\tau_{i}$  
will be $\sqrt{r}$ times smaller than the energy density of the plasma, i.e.
$B_{i} = \sqrt{ 4 \pi^2\, {\mathcal N} \,r\,/15} T_{i}^2$ where ${\mathcal N}$ denotes the effective number of spin degrees of freedom of the radiation 
plasma\footnote{The ratio between the magnetic energy density and the energy density of the flow is
$r = B^2/(8 \pi \rho)$ where in a radiation plasma (i.e. $p = \rho/3$ in Eq. (\ref{pmexplicit0})) we have $\rho= {\mathcal N} \pi^2 T^4/30$ and $T(\tau) = T_{i} (\tau/\tau_{i})^{-1/3}$.} and the conductivity $\sigma_{c}$ scales linearly with the temperature;  the diffusivity 
increases as a function of the proper time i.e. $\eta(\tau) = \eta_{i} (\tau/\tau_{i})^{\gamma}$ with $0< \gamma < 1$ ($\gamma=1/3$ for the radiation plasma). But this means that for the amplification of the hypermagnetic field the dynamics of $\psi$ is less relevant than the initial value of the magnetic diffusivity. Suppose indeed that $W=0$ so that Eq. (\ref{pmexplicit0}) implies $\dot{\psi} = \psi_{i}/\tau$; then  Eq. (\ref{exsol2}) gives
\begin{equation}
Q(\tau) = 1 + \frac{\eta_{i}}{\lambda\tau_{i}} \biggl[ 1 - \biggl(\frac{\tau_{i}}{\tau}\biggr)^{\lambda} \biggr], \qquad P(\tau) = - \alpha_{B}\biggl( \frac{\psi_{i}}{ \lambda M} \biggr) \biggl(\frac{\eta_{i}}{\tau_{i}}\biggr) \biggl[ 1 - \biggl(\frac{\tau_{i}}{\tau}\biggr)^{\lambda} \biggr],
\label{exsol3}
\end{equation}
where $\lambda = 1 - \gamma$ ($\lambda = 2/3$ for the radiation plasma).
More complicated evolutions of $\psi$ do not affect the scaling of the solution but can suppress the field even further. For instance if $\ddot{\psi} \ll \dot{\psi}/\tau$ then $\dot{\psi} \simeq -\tau W_{,\varphi}$; in this case $\psi$ is approximately constant and $P \to 0$. In the opposite limit (i.e. when $\psi$ oscillates) $Q(\tau)$ and $P(\tau)$ will have a  different analytical expression\footnote{ This aspect can be explicitly verified in the case of
massive potential (i.e. $W= m^2 \psi^2/2$) where the equation for $\psi$ (see Eq. (\ref{pmexplicit0})) can be solved exactly in terms of the Bessel functions 
$J_{\nu}(x)$ and $Y_{\nu}(x)$ with $\nu=0$ and argument $m\tau$.} but the solution itself will scale in the same way as a function of the initial 
value of the diffusivity $\eta$. This is sufficient to confirm that the results derived by approximating (and solving) the governing equations coincide with the limit of the exact solutions.

In summary the evolution of gauge fields in a relativistic plasma containing simultaneously vector and axial vector currents can be consistently formulated in a generally covariant framework that is relevant both for the electroweak epoch and for the hydromagnetic models of multiparticle dynamics. After obtaining the anomalous hypermagnetic diffusivity equation we demonstrated how the perfectly conducting limit washes out the magnetic and the vortical currents.  To amplify the hypermagnetic energy density, the flow and the anomalous currents must obey a set of necessary (but not sufficient) conditions extending to the relativistic domain the conventional notion of Beltrami field. These generalized constraints are satisfied, in particular, by an irrotational and boost-invariant flow with vanishing four-acceleration.

\end{document}